%% file: pip_gpd_paper.tex
\documentclass[preprint,12pt]{elsarticle}  

\usepackage{graphicx}  
\usepackage{dcolumn}   
\usepackage{bm}        
\usepackage{amssymb}   
\usepackage{xcolor}    
\usepackage{lipsum}
\usepackage{mathtools}
\journal{Physics Letters B}

\begin{document}

\begin{frontmatter}

\title{A multidimensional study of the structure function ratio $\sigma_{LT'}/\sigma_{0}$ from hard exclusive $\pi^+$ electro-production off protons in the GPD regime}

\newcommand*{\ANL}{Argonne National Laboratory, Argonne, Illinois 60439}
\newcommand*{\ANLindex}{1}
\newcommand*{\CSUDH}{California State University, Dominguez Hills, Carson, CA 90747}
\newcommand*{\CSUDHindex}{2}
\newcommand*{\CANISIUS}{Canisius College, Buffalo, NY}
\newcommand*{\CANISIUSindex}{3}
\newcommand*{\SACLAY}{IRFU, CEA, Universit\'{e} Paris-Saclay, F-91191 Gif-sur-Yvette, France}
\newcommand*{\SACLAYindex}{4}
\newcommand*{\CNU}{Christopher Newport University, Newport News, Virginia 23606}
\newcommand*{\CNUindex}{5}
\newcommand*{\UCONN}{University of Connecticut, Storrs, Connecticut 06269}
\newcommand*{\UCONNindex}{6}
\newcommand*{\DUKE}{Duke University, Durham, North Carolina 27708-0305}
\newcommand*{\DUKEindex}{7}
\newcommand*{\DUQUE}{Duquesne University, Pittsburgh PA, 15282, USA}
\newcommand*{\DUQUEindex}{8}
\newcommand*{\FU}{Fairfield University, Fairfield CT 06824}
\newcommand*{\FUindex}{9}
\newcommand*{\FERRARAU}{Universita' di Ferrara , 44121 Ferrara, Italy}
\newcommand*{\FERRARAUindex}{10}
\newcommand*{\FIU}{Florida International University, Miami, Florida 33199}
\newcommand*{\FIUindex}{11}
\newcommand*{\FSU}{Florida State University, Tallahassee, Florida 32306}
\newcommand*{\FSUindex}{12}
\newcommand*{\GWUI}{The George Washington University, Washington, DC 20052}
\newcommand*{\GWUIindex}{13}
\newcommand*{\GSIFFN}{GSI Helmholtzzentrum fur Schwerionenforschung GmbH, D-64291 Darmstadt, Germany}
\newcommand*{\GSIFFNindex}{14}
\newcommand*{\INFNFE}{INFN, Sezione di Ferrara, 44100 Ferrara, Italy}
\newcommand*{\INFNFEindex}{15}
\newcommand*{\INFNFR}{INFN, Laboratori Nazionali di Frascati, 00044 Frascati, Italy}
\newcommand*{\INFNFRindex}{16}
\newcommand*{\INFNGE}{INFN, Sezione di Genova, 16146 Genova, Italy}
\newcommand*{\INFNGEindex}{17}
\newcommand*{\INFNRO}{INFN, Sezione di Roma Tor Vergata, 00133 Rome, Italy}
\newcommand*{\INFNROindex}{18}
\newcommand*{\INFNTUR}{INFN, Sezione di Torino, 10125 Torino, Italy}
\newcommand*{\INFNTURindex}{19}
\newcommand*{\INFNPAV}{INFN, Sezione di Pavia, 27100 Pavia, Italy}
\newcommand*{\INFNPAVindex}{20}
\newcommand*{\ORSAY}{Universit'{e} Paris-Saclay, CNRS/IN2P3, IJCLab, 91405 Orsay, France}
\newcommand*{\ORSAYindex}{21}
\newcommand*{\Juelich}{Institute fur Kernphysik (Juelich), Juelich, Germany}
\newcommand*{\Juelichindex}{22}
\newcommand*{\JMU}{James Madison University, Harrisonburg, Virginia 22807}
\newcommand*{\JMUindex}{23}
\newcommand*{\KNU}{Kyungpook National University, Daegu 41566, Republic of Korea}
\newcommand*{\KNUindex}{24}
\newcommand*{\LAMAR}{Lamar University, 4400 MLK Blvd, PO Box 10046, Beaumont, Texas 77710}
\newcommand*{\LAMARindex}{25}
\newcommand*{\MIT}{Massachusetts Institute of Technology, Cambridge, Massachusetts  02139-4307}
\newcommand*{\MITindex}{26}
\newcommand*{\MISS}{Mississippi State University, Mississippi State, MS 39762-5167}
\newcommand*{\MISSindex}{27}
\newcommand*{\ITEP}{National Research Centre Kurchatov Institute - ITEP, Moscow, 117259, Russia}
\newcommand*{\ITEPindex}{28}
\newcommand*{\UNH}{University of New Hampshire, Durham, New Hampshire 03824-3568}
\newcommand*{\UNHindex}{29}
\newcommand*{\NMSU}{New Mexico State University, PO Box 30001, Las Cruces, NM 88003, USA}
\newcommand*{\NMSUindex}{30}
\newcommand*{\NSU}{Norfolk State University, Norfolk, Virginia 23504}
\newcommand*{\NSUindex}{31}
\newcommand*{\OHIOU}{Ohio University, Athens, Ohio  45701}
\newcommand*{\OHIOUindex}{32}
\newcommand*{\ODU}{Old Dominion University, Norfolk, Virginia 23529}
\newcommand*{\ODUindex}{33}
\newcommand*{\JLUGiessen}{II Physikalisches Institut der Universitaet Giessen, 35392 Giessen, Germany}
\newcommand*{\JLUGiessenindex}{34}
\newcommand*{\URICH}{University of Richmond, Richmond, Virginia 23173}
\newcommand*{\URICHindex}{35}
\newcommand*{\ROMAII}{Universita' di Roma Tor Vergata, 00133 Rome Italy}
\newcommand*{\ROMAIIindex}{36}
\newcommand*{\MSU}{Skobeltsyn Institute of Nuclear Physics, Lomonosov Moscow State University, 119234 Moscow, Russia}
\newcommand*{\MSUindex}{37}
\newcommand*{\SCAROLINA}{University of South Carolina, Columbia, South Carolina 29208}
\newcommand*{\SCAROLINAindex}{38}
\newcommand*{\TEMPLE}{Temple University,  Philadelphia, PA 19122 }
\newcommand*{\TEMPLEindex}{39}
\newcommand*{\JLAB}{Thomas Jefferson National Accelerator Facility, Newport News, Virginia 23606}
\newcommand*{\JLABindex}{40}
\newcommand*{\UTFSM}{Universidad T\'{e}cnica Federico Santa Mar\'{i}a, Casilla 110-V Valpara\'{i}so, Chile}
\newcommand*{\UTFSMindex}{41}
\newcommand*{\INSUBRIA}{Universit\`{a} degli Studi dell'Insubria, 22100 Como, Italy}
\newcommand*{\INSUBRIAindex}{42}
\newcommand*{\BRESCIA}{Universit`{a} degli Studi di Brescia, 25123 Brescia, Italy}
\newcommand*{\BRESCIAindex}{43}
\newcommand*{\UCR}{University of California Riverside, 900 University Avenue, Riverside, CA 92521, USA}
\newcommand*{\UCRindex}{44}
\newcommand*{\GLASGOW}{University of Glasgow, Glasgow G12 8QQ, United Kingdom}
\newcommand*{\GLASGOWindex}{45}
\newcommand*{\YORK}{University of York, York YO10 5DD, United Kingdom}
\newcommand*{\YORKindex}{46}
\newcommand*{\VIRGINIA}{University of Virginia, Charlottesville, Virginia 22901}
\newcommand*{\VIRGINIAindex}{47}
\newcommand*{\WM}{College of William and Mary, Williamsburg, Virginia 23187-8795}
\newcommand*{\WMindex}{48}
\newcommand*{\YEREVAN}{Yerevan Physics Institute, 375036 Yerevan, Armenia}
\newcommand*{\YEREVANindex}{49}

\newcommand*{\NOWANL}{Argonne National Laboratory, Argonne, Illinois 60439}
\newcommand*{\NOWJLAB}{Thomas Jefferson National Accelerator Facility, Newport News, Virginia 23606}

\author[toJLUGiessen,toUCONN]{S.~Diehl}
\author[toUCONN]{A.~Kim}
\author[toUCONN]{K.~Joo}

\author[toJLAB]{P.~Achenbach}
\author[toVIRGINIA,toFSU]{Z.~Akbar}
\author[toODU]{M.J.~Amaryan}
\author[toTEMPLE]{H.~Atac}
\author[toJLAB]{H.~Avagyan}
\author[toWM]{C.~Ayerbe Gayoso}
\author[toFIU]{L.~Baashen}

% ...
\address[toUCONN]{\UCONN} 
% ...
\address[toFIU]{\FIU} 
 \address[toFSU]{\FSU} 
% ...
\address[toODU]{\ODU} 
 \address[toJLUGiessen]{\JLUGiessen} 
% ...
\address[toTEMPLE]{\TEMPLE} 
 \address[toJLAB]{\JLAB} 
% ...
\address[toVIRGINIA]{\VIRGINIA} 
 \address[toWM]{\WM} 
% ...

 \fntext[toNOWANL]{Current address: Argonne, Illinois 60439 }
 \fntext[toNOWJLAB]{Current address: Newport News, Virginia 23606 }

\begin{abstract}
A multidimensional extraction of the structure function ratio $\sigma_{LT'}/\sigma_{0}$ from the hard exclusive $\vec{e} p \to e^\prime n \pi^+$ reaction above the resonance region has been performed. The study was done based on beam-spin asymmetry measurements using a 10.6~GeV incident electron beam on a liquid-hydrogen target and the CLAS12 spectrometer at Jefferson Lab. The measurements focus on the very forward regime ($t/Q^{2}$ $\ll$ 1) with a wide kinematic range of $x_{B}$ in the valence regime (0.17 $<$ $x_{B}$ $<$ 0.55), and virtualities $Q^{2}$ ranging from 1.5 GeV$^{2}$ up to 6 GeV$^{2}$. The results and their comparison to theoretical models based on Generalized Parton Distributions demonstrate the sensitivity to chiral-odd GPDs and the directly related tensor charge of the nucleon. In addition, the data is compared to an extension of a Regge formalism at high photon virtualities. It was found that the Regge model provides a better description at low $Q^{2}$, while the GPD model is more appropriate at high $Q^{2}$.

\end{abstract}

\begin{keyword}
Exclusive single pion, Eletroproduction, GPD, CLAS12
\PACS{13.60.Le, 14.20.Dh, 14.40.Be, 24.85.+p}
\end{keyword}

\end{frontmatter}

\begin{figure}[!ht]
\begin{center}
    \includegraphics[width=0.4\textwidth]{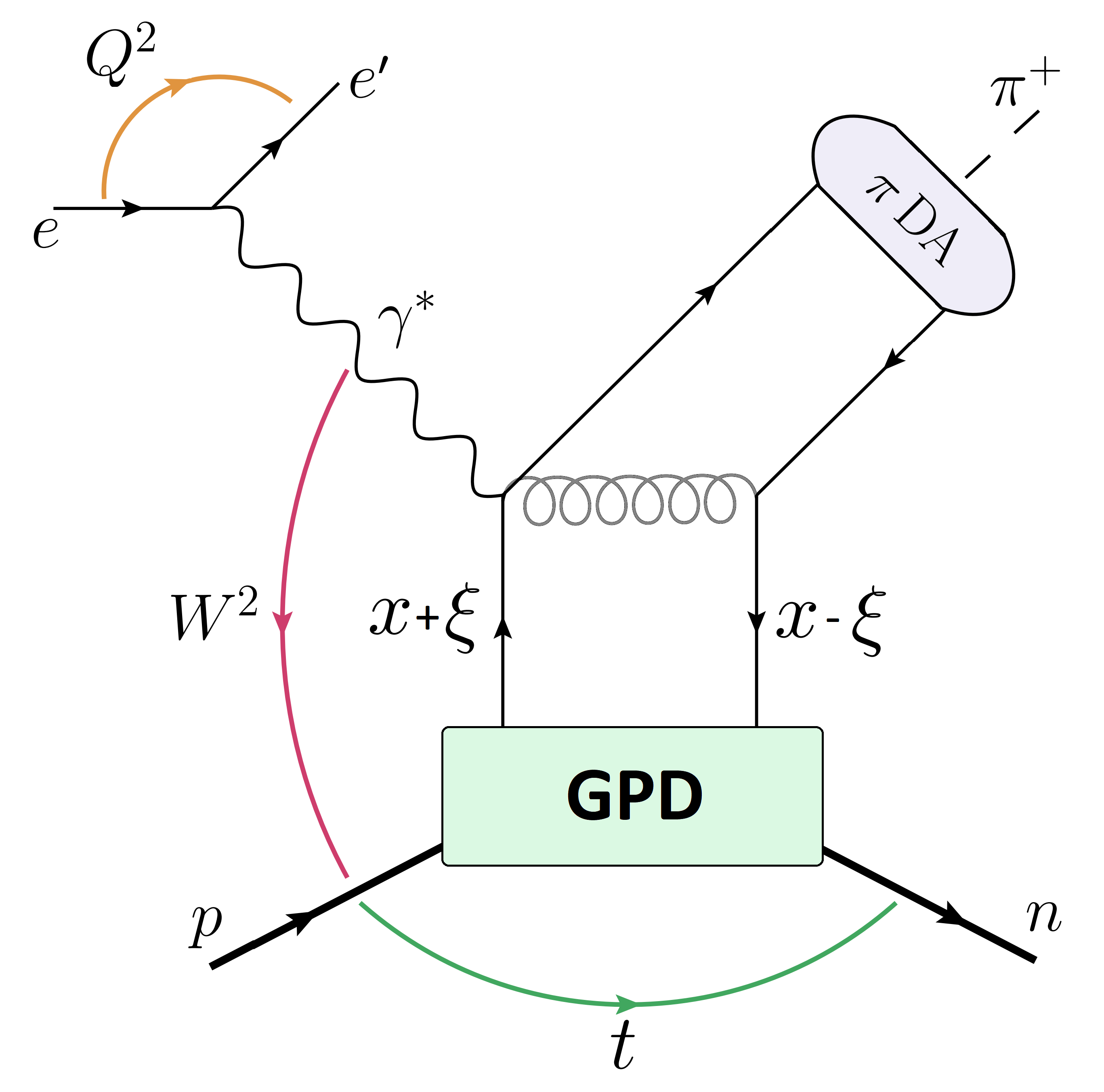} 
	\caption{Hard exclusive electro-production of a pion on the proton in very forward kinematics ($-t/Q^2 \ll 1$), described by GPDs~\cite{previous1, previous2}.}
	\label{fig:production_mechanism}
	\end{center}
\end{figure}

Generalized Parton Distributions (GPDs) \cite{Rad97-5, Col97-6, Brod94-7} provide direct access to the three-dimensional structure of the nucleon by correlating information about the transverse position and the longitudinal momentum of the quarks and gluons within the nucleon. Besides deeply virtual Compton scattering (DVCS), also deeply virtual meson production (DVMP) can be used to access GPDs.
The factorization of the DVMP process into a perturbatively calculable hard-scattering part and two soft hadronic matrix elements, parameterized by GPDs and a meson distribution amplitude (DA) as shown in Fig.~\ref{fig:production_mechanism} has been proven for longitudinally polarized virtual photons at large photon virtuality $Q^{2}$, large energy $W$ and fixed Bjorken-x \cite{fact1,fact2}. The contribution of transversely polarized virtual photons for which factorisation is not explicitly proven, is typically treated as a higher twist-effect in current phenomenological models \cite{previous1}.

Previous experimental results based on the hard exclusive scattering of 27.6 GeV/c un-polarized electron and positron beams off polarized and un-polarized hydrogen targets at HERMES \cite{ HERMES02, HERMES08, HERMES10}, up to 6 GeV/c polarized and un-polarized electron beams at CLAS \cite{DeMasi2008, Bedlinskiy2012, Bedlinskiy2014, Bedlinskiy2017, Kim2017, Bosted_pi02017, Bosted_piplus2017, Zhao2019}
and hall A at JLAB \cite{hallA_2012, hallA_2016, hallA_2017} and based on 160 GeV/c polarized muon beams at COMPASS \cite{COMPASS20}, as well as theoretical studies \cite{previous1, previous2, DK07, DMP17, SS19} of hard exclusive pseudoscalar meson electro-production, especially $\pi^0$ and $\eta$ electroproduction~\cite{Bedlinskiy2014, Bedlinskiy2017, Kim2017, Zhao2019, previous1, previous2, Ahmad2009, Goldstein2011}, have shown that the asymptotic leading-twist approximation is not sufficient to describe the experimental results from the existing measurements.
It was found that there are strong contributions from transversely polarized virtual photons that have to be considered by including contributions from chiral-odd GPDs ($H_{T}$, $\widetilde{H}_{T}$, $E_{T}$, and $\widetilde{E}_{T}$) in addition to the chiral-even GPDs ($H$, $\widetilde{H}$, $E$ and $\widetilde{E}$), which depend on the momentum fraction of the parton $x$, the skewness $\xi$ and the four-momentum transfer to the nucleon $t$.

While chiral-even GPDs can be related to the well known nucleon form factors \cite{Liutti}, only a few phenomenological constraints exist for the chiral-odd GPDs. For example, the first moment of $2 \widetilde{H}_{T} + E_{T}$ can be interpreted as the proton’s transverse anomalous magnetic moment \cite{Burk06}, while in the forward limit, $H_{T}$ becomes the transversity structure function $h_{1}$, which is directly related to the still unknown tensor charge of the nucleon \cite{Liutti}. 

In exclusive meson production experiments, GPDs are typically accessed through differential cross sections and beam and target polarization asymmetries \cite{Dre1992, Are1997, Die2005}. The focus of this work is on the extraction of the structure function ratio  $\sigma_{LT'}/\sigma_{0}$ from beam-spin asymmetry measurements.
In the one-photon exchange approximation the beam-spin asymmetry is defined as \cite{Dre1992, Are1997}:
\begin{eqnarray}\label{eq:BSA}
	BSA  = \frac{\sqrt{2 \epsilon (1 - \epsilon)}  \frac{\sigma_{LT^{\prime}}}
	{\sigma_{0}}\sin\phi}
	{1 + \sqrt{2 \epsilon (1 + \epsilon)}\frac{\sigma_{LT} }{\sigma_{0}} \cos\phi
	+ \epsilon \frac{\sigma_{TT}}{\sigma_{0}}  \cos2\phi},
\end{eqnarray}
where the structure functions $\sigma_{L}$ and $\sigma_{T}$, which contribute to $\sigma_{0} = \sigma_{T} + \epsilon \sigma_{L}$, correspond to coupling to longitudinal and transverse virtual photons, and $\epsilon$ describes the flux ratio of longitudinally and transversely polarized virtual photons. $\sigma_{LT}$, $\sigma_{TT}$ and the polarized structure function $\sigma_{LT^\prime}$ describe the interference between their amplitudes. $\phi$ is the azimuthal angle between the electron scattering plane and the hadronic reaction plane. 

$\sigma_{LT^\prime}$ can be expressed through the convolutions of GPDs with sub-process amplitudes (see Eq. \ref{eqn:Htdep}) and contains the products of chiral-odd and chiral-even terms \cite{previous1}. For the $\pi^{+}$ channel, the imaginary parts of chiral-odd GPDs in $\sigma_{LT^\prime}$ are significantly amplified by the pion pole term, where the contributions of GPDs are largely imaginary and those of the pion pole are real and can be accurately calculated. 
Due to this feature, polarized $\pi^{+}$ observables show an increased sensitivity to chiral-odd GPDs like $H_{T}$ and can therefore be used to probe fundamental observables like the tensor charge $\delta_{T}$ for up ($u$) and down ($d$) quarks of the nucleon by
\begin{eqnarray}
    \delta_{T}^{u,d} = \int^{1}_{\xi - 1} dx H^{u,d}_{T}(x, \xi, t=0),
\end{eqnarray}
with the longitudinal momentum transfer $\xi$ \cite{Ahmad2009}.
Due to the missing pion pole contribution, this sensitivity is much lower for exclusive $\pi^{0}$ and $\eta$ production. In addition, $\pi^{+}$ observables are especially suited to access $H_{T}$, in contrast to $\pi^{0}$ and $\eta$ production, due to the flavour composition of the charged pions.

An alternative description of hard excluisve pion production is based on Regge models.
In these models, the interaction is mediated by the exchange of trajectories in the $t$ channel. While Regge models were initially extensively studied for photoproduction ($Q^{2}$ = 0) \cite{GLV97}, an extension to the deeply virtual regime has been implemented within the Laget model (JML), which is based on Reggeized $\pi^{+}$ and $\rho^{+}$ meson exchanges in the $t$-channel \cite{JML20prog, JML20} and unitarity cuts \cite{JML10, JML11}. The $t$-channel exchange of the pion and the $\rho$ rely on the canonical VGL \cite{VGL98} description, supplemented by the $t$-dependent electromagnetic form factor introduced in Ref. \cite{JML04}. Alone these pole terms lead to a vanishing BSA. The elastic $\pi-N$ \cite{JML11} and inelastic $\rho-N$ unitarity cuts \cite{JML10, JML11} provide the phase necessary to get a non-zero BSA, through their interference with the Regge poles. The JML model, which provides a unified description at the real photon point, as well as in the virtual photon sector, nicely reproduces the recent CLAS \cite{PGG13} and HERMES \cite{AHC08} data on un-polarized $\pi^{+}$ electro-production cross sections.

Altogether, two theoretical descriptions are available for hard exclusive $\pi^{+}$ electro-production. While the JML model starts at the real photon point and extends to the deeply virtual regime, a firm QCD foundation exists for the GPD model within the Bjorken regime and its applicability must be tested in the accessible $Q^{2}$ range.

Previous measurements of the hard exclusive $\pi^{+}$ production BSA ({\it i.e.} \cite{Diehl20}) only provided a binning in $-t$ and $\phi$, while the virtuality $Q^{2}$ and the Bjorken scaling variable $x_{B}$ where integrated over the complete accessible range due to limited statistics.
In addition, only a limited range in $Q^{2}$ could be accessed due to the low electron beam energies that were available for these studies.
For a precise comparison to theoretical models and especially for a study of higher-twist effects, a multidimensional study in $t$, $\phi$, $x_{B}$ and $Q^{2}$ with fine binning is needed to reduce uncertainties and to access the kinematic dependencies of the involved GPDs. In addition, a fully multidimensional study can provide a better comparison between the theoretical models and the data and help to investigate the validity of the two models.


For the present study, hard exclusive $\pi^+$ electro-production was measured at Jefferson Lab with CLAS12 (CEBAF Large Acceptance Spectrometer for operation at 12 GeV) \cite{VDB20}. Beam-spin asymmetries in forward kinematics were extracted over a wide range in $Q^2$, $x_{B}$ and $\phi$. The incident electron beam was longitudinally polarized and had an energy of 10.6~GeV and an average current of 40-55~nA, impinging on a 5-cm-long un-polarized liquid-hydrogen target placed at the center of the solenoid magnet of CLAS12. The CLAS12 forward detector consists of six identical sectors within a toroidal magnetic field. The momentum and the charge of the particles were determined by 3 regions of drift chambers from the curvature of the particle trajectories in the magnetic field. The electron identification was based on a lead-scintillator electromagnetic sampling calorimeter in combination with a Cherenkov counter. Positive pions were identified by time-of-flight measurements. 
Based on the high statistics of CLAS12, a precise, multidimensional study of the cross section ratio  $\sigma_{LT'}/\sigma_{0}$ becomes possible for the first time.

\begin{figure}[!ht]
	\centering
		\includegraphics[width=0.45\textwidth]{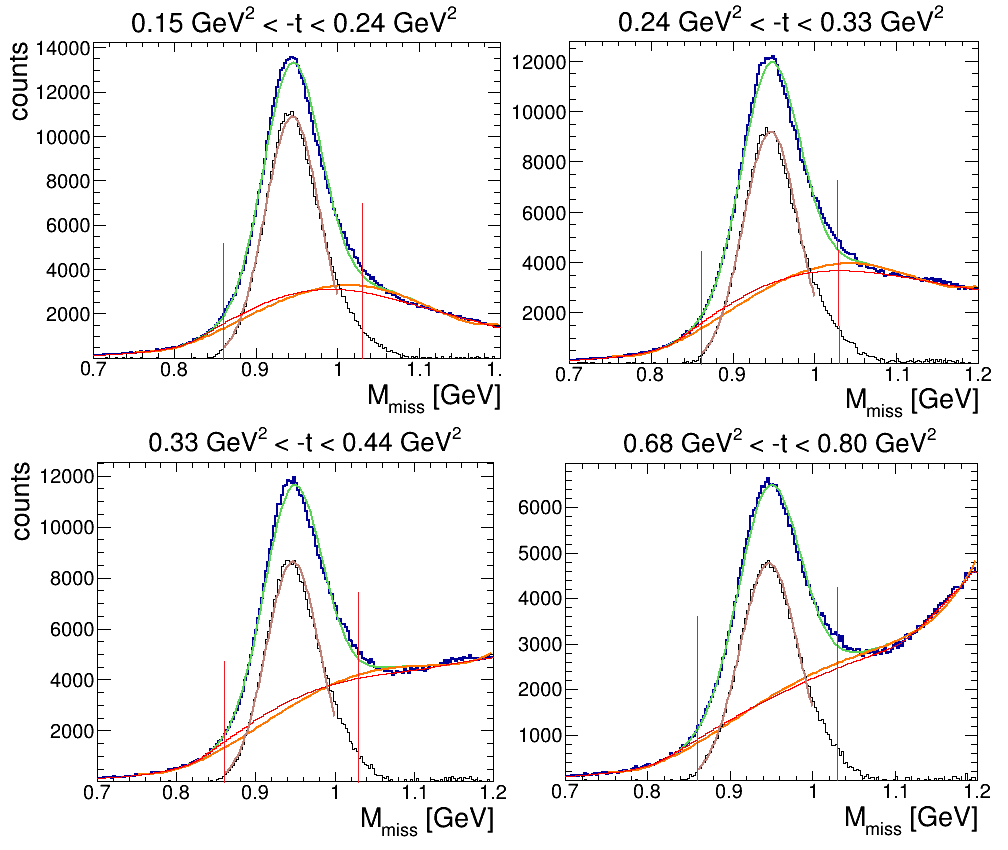}
	\caption{Missing mass spectrum of $e'\pi^{+}X$ in the region of the missing neutron peak for  selected bins of $-t$ in the forward region. The raw distributions (upper histogram in each plot) were fit with a Gaussian (green curve) and a polynomial background (orange curve). For comparison, the background histogram obtained with the CERN-ROOT based background estimator applying a sensitive nonlinear iterative peak clipping algorithm \cite{Morh97} is shown in red and the background subtracted missing neutron peak is displayed as a black histogram fitted with a Gaussian (brown). The cut borders for the event selection are shown as vertical lines.}
	\label{fig:missing_mass}
\end{figure}

For the selection of deeply inelastic scattered electrons, cuts on $Q^{2}~>$ 1.5~GeV$^{2}$, $y < 0.75$ and on the invariant mass of the hadronic final state $W~>~2$~GeV, were applied. To select the exclusive $e^{\prime} \pi^{+} n$ final state, events with exactly one electron and one $\pi^{+}$ were detected, and the missing neutron was selected via a cut on the neutron peak in the $e'\pi^{+}X$ missing mass spectrum. Figure \ref{fig:missing_mass} shows the missing mass spectrum for $e'\pi^{+}X$ in the region around the missing neutron peak for selected bins of $-t$ in the forward region, integrated over $Q^{2}$ and $x_{B}$.

As illustrated in Fig. \ref{fig:missing_mass}, the signal-to-background ratio decreases with $-t$ from $\approx$~4.5 at $-t$ close to the threshold $t_{min}$ to $\approx$~2 for $-t \approx t_{min} +1$~GeV$^2$, making a background subtraction necessary for beam-spin asymmetry extractions.
The observed background behaviour was found to be nearly independent of the $Q^{2}$ and $x_{B}$ bin.
To determine the signal and background counts, the complete distribution (signal + background) was fit with a Gaussian (describing the signal) plus a third-order polynomial (describing the background). After the combined fit, the signal and background contributions can be separated and integrated within a 2 $\sigma$ region of the Gaussian distribution.
As a crosscheck, another background histogram was obtained with the CERN-root based background estimator applying a sensitive nonlinear iterative peak clipping algorithm \cite{Morh97}. The obtained background was found to be very similar to the result from a full fit of the signal and background function (see Fig. \ref{fig:missing_mass}), and was used to estimate the systematic uncertainty of the background subtraction.

\begin{figure}[!ht]
	\centering
		\includegraphics[width=0.45\textwidth]{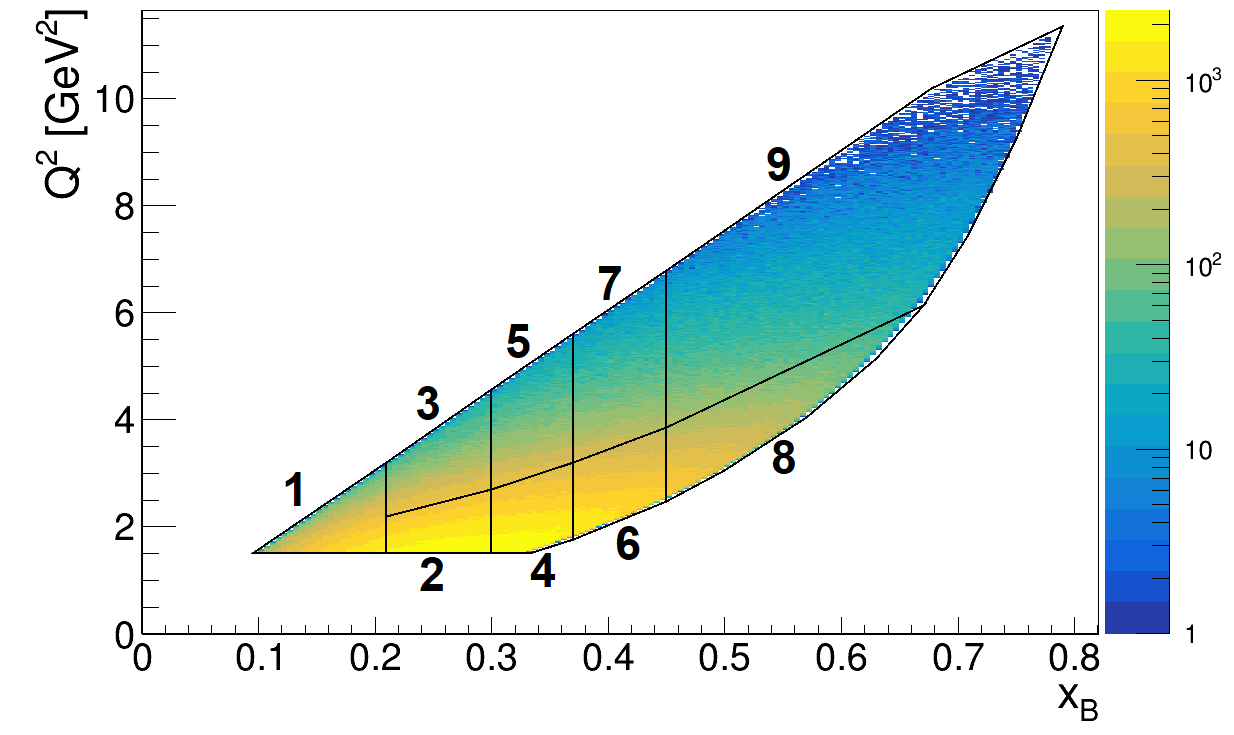}
	\caption{Distribution of $Q^{2}$ versus $x_{B}$. The bin boundaries are shown as black lines and the bin numbering is given. The bin borders are also provided in the supplemental material \cite{supl}.}
	\label{fig:q2x_bins}
\end{figure}

Figure \ref{fig:q2x_bins} shows the $Q^{2}$ versus $x_{B}$ distribution of the exclusive events, together with the binning scheme applied for the multidimensional study.
For each of the nine  $Q^{2}-x_{B}$ bins, up to six bins in $-t$ and 12 bins in $\phi$ were defined to extract the beam-spin asymmetry (BSA).

The BSA was determined experimentally from the number of counts with positive and negative helicity ($N^{\pm}_{i}$), in a specific bin $i$ as:
\begin{eqnarray}
	BSA_{i} = \frac{1}{P_{e}} \frac{N^{+}_{i} - N^{-}_{i}}{N^{+}_{i} + N^{-}_{i}},
\end{eqnarray}
where $P_{e}$ is the average magnitude of the beam polarization. $P_{e}$ was measured with a M{\o}ller polarimeter upstream of CLAS12 to be 86.3\%$\pm$2.6\%. 
To obtain the signal counts, a full fit of the signal and background as described above was applied for each multidimensional bin in $Q^{2}$, $x_{B}$, $-t$ and $\phi$ and for each helicity state separately. The number of counts and their uncertainty were then given by the integral over the fit function of the signal distribution and the uncertainty of the beam-spin asymmetry was calculated based on standard error propagation.

\begin{figure}[!ht]
	\centering
		\includegraphics[width=0.45\textwidth]{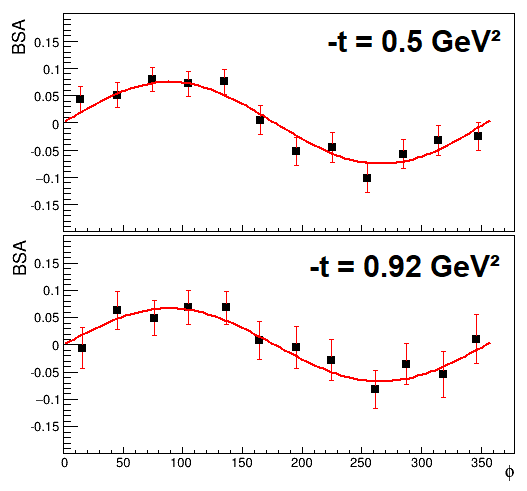}
	\caption{Beam-spin asymmetry as a function of $\phi$ for representative $-t$ bins of $Q^{2}-x_{B}$ bin 9 ($Q^{2}$ = 5.8 GeV$^{2}$, $x_{B}$ = 0.55). The vertical error bars show the statistical uncertainty of each point. The red line shows the fit with the functional form of Eq. (\ref{eq:BSA}).}
	\label{fig:sinphi_fit}
\end{figure}

To extract the structure function ratio $\sigma_{LT'}/\sigma_{0}$, the beam-spin asymmetry was plotted as a function of the azimuthal angle $\phi$. Then a fit of the data with a $\sin\phi$ function was applied. The flux ratio $\epsilon$ (see Eq.\ref{eq:BSA}) was calculated for each bin based on the electron kinematics. Figure \ref{fig:sinphi_fit} shows the beam-spin asymmetry as a function of $\phi$ in two different $-t$ bins for the example of $Q^{2}-x_{B}$ bin 9.
Even in the highest $Q^{2}$ bin shown, a precise measurement of the $\phi$ dependence is possible. As expected, the $\phi$-dependence can be well described by the assumed $\sin\phi$ shape. The impact of the denominator terms in Eq.\ref{eq:BSA} on $\sigma_{LT'}/\sigma_{0}$ was studied during the analysis using different extraction methods and was found to be on average 2.7\% and, therefore, much smaller than the statistical and the total systematic uncertainty, and was considered as a systematic uncertainty.

The main source of systematic uncertainty is given by the background subtraction. It was evaluated by comparing the two described background subtraction methods. The variation between the two methods which was in average 4.9\% is considered as systematic uncertainty.
The systematic effect due to the uncertainty of the beam polarization (3.4\%) was determined based on the uncertainty of the measurement with the M{\o}ller polarimeter.
To estimate the impact of acceptance and bin-migration effects, a realistic Monte Carlo simulation including all detector effects was performed. The impact of these effects was evaluated by comparing the injected and reconstructed asymmetries and was found to be in the order of 3.6\%. Systematic uncertainties due to radiative effects have been studied based on Ref. \cite{AAB02}, and were found to be in the order of 3.0\%. 
Several additional sources of systematic uncertainty, including particle identification and the effect of fiducial volume definitions, were investigated and found to give a small contribution to the total systematic uncertainty ($<$1.5\%).
The total systematic uncertainty in each bin is defined as the square-root of the quadratic sum of the uncertainties from all sources. On average it was found to be on the order of 8.3\%, which is smaller than the statistical uncertainty in most kinematic bins.

\begin{figure*}[!ht]
	\centering
		\includegraphics[width=1.0\textwidth]{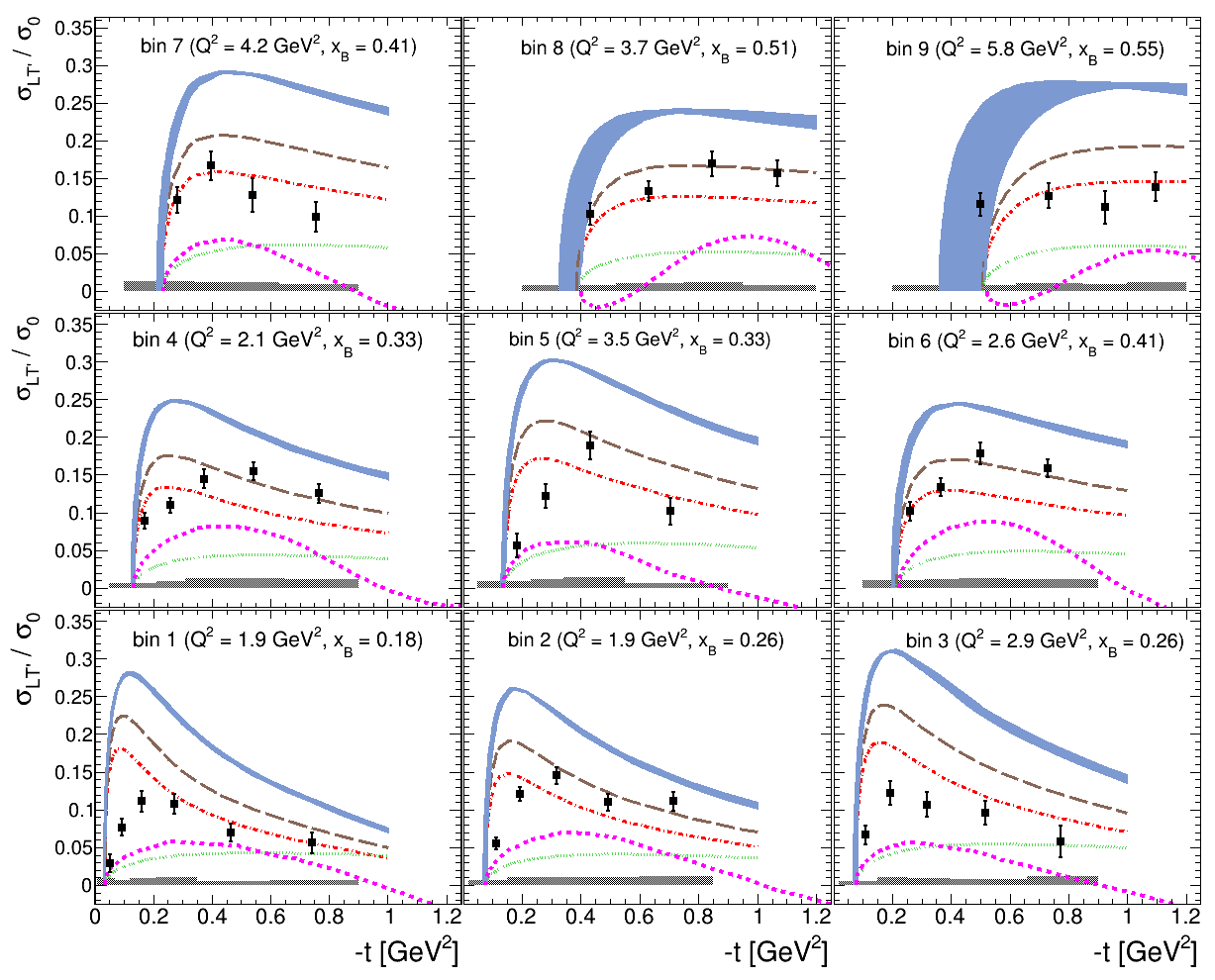}
	\caption{$\sigma_{LT'}/\sigma_{0}$ and its statistical uncertainty as a function of $-t$ in the forward kinematic regime and its systematic uncertainty (grey bins). The bold dotted magenta line shows the theoretical prediction from the Regge based JML model \cite{JML20prog}. The blue band shows the theoretical prediction from the GPD-based GK model \cite{GK09,BBC18}. The dashed brown and the dash-dotted red curve show the effect of increasing the GPD $H_{T}$ by an overall factor of 1.5 and 2.0 for the mean kinematics.The dotted green curve shows the theory result under the assumption that no pion pole term is contributing.  The corresponding result tables can be found in the supplemental material \cite{supl} and can be downloaded from Ref. \cite{CLASdata}.}
	\label{fig:ALU_theory}
\end{figure*}

Figure \ref{fig:ALU_theory} shows the final results for $\sigma_{LT'}/\sigma_{0}$ in the region of $-t$ up to 0.8~GeV$^{2}$ - 1.2~GeV$^{2}$, depending on the $Q^{2}$ bin ($-t/Q^{2} \approx 0.2 - 0.4$), where the leading-twist GPD framework is applicable and compares them 
to the theoretical predictions from the JML model \cite{JML20prog}, which is based on hadronic degrees of freedom and to the predictions from the GPD-based model by Goloskokov and Kroll (GK) \cite{GK09,BBC18}. The band on the theoretical prediction represents the variation of the mean value of $Q^{2}$ and $x_{B}$ within each multidimensional bin.
The increasing width of these bands for bins 8 and 9, which cover a larger $x_{B}$ and $Q^{2}$ range than the other bins, clearly shows the advantages of a fine multidimensional binning for a precise theory comparison.
The structure function ratio $\sigma_{LT'}/\sigma_{0}$ is clearly positive in all kinematic bins and shows a typical shape that can be explained by the contributing structure functions.
The non-$\phi$-dependent cross section $\sigma_{0} = \sigma_{T} + \epsilon \sigma_{L}$ is typically forward peaked due to the pion pole term contribution, while $\sigma_{LT^\prime}$ is constrained to be zero at $t=t_{min}$ due to angular momentum conservation. 

The GK model includes chiral-odd GPDs to calculate the contributions from the transversely polarized virtual photon amplitudes, with their $t$-dependence incorporated from Regge phenomenology. The GPDs are constructed from double distributions and constrained by the latest results from lattice QCD and transversity parton distribution functions \cite{GK09}. A special emphasis is given to the GPDs $H_{T}$ and $\overline{E}_{T} = 2 \widetilde{H}_T + E_T$, while contributions from other chiral-odd GPDs are neglected in the calculations, unlike chiral-even GPDs. The pion pole contribution to the amplitudes is taken into account for longitudinally and transversely polarized virtual photons. 

$\sigma_{LT^\prime}$ can be expressed through the convolutions of GPDs with sub-process amplitudes (twist-2 for the longitudinal and twist-3 for the transverse amplitudes) and contains the products of chiral-odd and chiral-even terms \cite{previous1}:
\begin{eqnarray}
	\sigma_{LT^\prime} \sim \xi \sqrt{1-\xi^{2}} \frac{\sqrt{-t'}}{2m} Im[ \langle\overline{E}_{T-eff}\rangle^{*} \langle\widetilde{H}_{eff}\rangle \nonumber\\
	+ \langle H_{T-eff}\rangle^{*} \langle\widetilde{E}_{eff}\rangle],
\label{eqn:sigma_GPD}
\end{eqnarray}
where $m$ is the proton mass and the ``eff'' in the subscript describes the inclusion of the pion pole term, {\it i.e.}
\begin{eqnarray}
\langle\widetilde{E}_{eff}\rangle = \langle\widetilde{E}_{\text{non-pole}}\rangle + c \frac{\rho_{\pi}}{t-m^{2}_{\pi}}\\
\langle\widetilde{H}_{eff}\rangle = \langle\widetilde{H}\rangle + \frac{\xi^{2}}{1-\xi^{2}} \langle\widetilde{E}_{eff}\rangle
\label{eqn:pion_pole}
\end{eqnarray}
with a factor $c = m_{p} Q^{2}/\xi$, the residue $\rho_{\pi}$ and the pion mass $m_{\pi}$ \cite{GK09}. 

For $\pi^{+}$ the imaginary part of small chiral-odd GPDs in $\sigma_{LT^\prime}$ is significantly amplified by the pion pole term, which is real and theoretically well described.
The strength of this effect is illustrated in Fig. \ref{fig:ALU_theory}, which shows the comparison between the calculation with and without considering the pion pole (blue band vs green dotted line). 
Due to this feature, polarized $\pi^{+}$ observables show an increased sensitivity to chiral-odd GPDs in contrast to the exclusive $\pi^{0}$ and $\eta$ production where the pole contribution is not present.
The pion pole is well determined from cross section measurements with an uncertainty of less than 10\%. Therefore, it cannot explain the observed overestimation of the experimental result by the theoretical prediction.
\newline
The denominator terms of the structure function ratio $\sigma_L$ and $\sigma_T$ can be expressed by \cite{previous1}:
\begin{eqnarray}
	\sigma_{L} \sim (1- \xi^{2}) \left|\langle \widetilde{H}_{eff} \rangle\right|^{2} - 2 \xi^{2} Re\left[ \langle \widetilde{H}_{eff} \rangle^{*} \langle \widetilde{E}_{eff} \rangle\right] \nonumber\\ - \frac{t'}{4m^{2}} \xi^{2} \left|\langle \widetilde{E}_{eff} \rangle\right|^{2}
\end{eqnarray}
\begin{eqnarray}
	\sigma_{T} \sim (1- \xi^{2}) \left|\langle H_{T-eff} \rangle\right|^{2} - \frac{t'}{8m^{2}} \left|\langle \overline{E}_{T-eff} \rangle\right|^{2}.
\end{eqnarray}
Due to the quark flavour composition of the pions, $\pi^{+}$ production is typically dominated by $H_{T}$, while the contribution from $\overline{E}_{T}$ is significantly smaller. In contrast to this, neutral psuedoscalar-mesons like $\pi^{0}$ and $\eta$ show a significantly stronger contribution from $\overline{E}_{T}$, except at very small values of $-t$ where $H_{T}$ dominates. 
Since  chiral even GPDs are much better known than their chiral odd counterparts, the strongest uncertainty for the theoretical prediction is expected from the so far poorly known GPD $H_{T}$ for which the dependence on the measured structure function ratio is given in Eq.\ref{eqn:Htdep}.
\begin{equation}
\label{eqn:Htdep}
	\frac{\sigma_{LT'}}{\sigma_{0}} \sim \frac{Im\left[\langle H_{T-eff}\rangle^{*} \langle\widetilde{E}_{eff}\rangle]\right]}{\left|\langle H_{T-eff} \rangle\right|^{2} + \epsilon \sigma_{L}}.
\end{equation}
The comparison between the experimental results and the theoretical predictions shows that the magnitude of the GK model calculations is overestimated, while the $t$-dependence of the measured $\sigma_{LT'}/\sigma_{0}$ values is, especially if the variation with $Q^{2}$ and $x_{B}$ is considered, much better, but not perfectly reproduced. This discrepancy of the magnitude might be due to the interplay of the pion pole term with the poorly known chiral-odd GPD $H_{T}$. Based on Eq.\ref{eqn:Htdep} the results especially hint on an underestimation of $H_{T}$.
To show the sensitivity of  $\sigma_{LT'}/\sigma_{0}$ on the GPD $H_{T}$, Fig. \ref{fig:ALU_theory} also contains calculations under the assumption that the GPD $H_{T}$ is increased by an overall factor of 1.5 (brown dashed line) and by a factor of 2.0 (red dash-dotted line). Due to the amplification by the pion pole term, a strong sensitivity to such a variation can be observed. After the modification of the GPD $H_{T}$, a significantly better agreement between the theoretical predictions and the experimental result is observed.

However, a change of $H_{T}$ will help as far as $\sigma_{LT'}/\sigma_{0}$ is concerned, but the consequences for other observables remain to be checked.
Especially observables with transversely polarized targets like the $\sin\phi_{S}$ modulation of the $A_{UT}$ moment for hard exclusive $\pi^{+}$ production, for which measurements based on HERMES data exist \cite{GK09} and various modulations of $A_{UT}$ and $A_{LT}$ for $\rho^{0}$ production \cite{GK14} show strong contributions from the transversity GPDs and need to be considered for the determination of $H_{T}$.
Altogether, a new global fit of the GPDs to all existing data, {\it e.g.} \cite{DeMasi2008, Zhao2019, Bedlinskiy2012, Bedlinskiy2014, Bedlinskiy2017}, as well as the aforementioned HERMES results and additional upcoming CLAS12 results on other mesons becomes necessary.
Here, the new multidimensional, high precision $\pi^{+}$ beam-spin asymmetry data from this work and its high sensitivity to the GPD $H_{T}$ due to the amplification by the pion pole, will allow a much better determination of this so far poorly known GPD. 
Based on the improvements in the knowledge of $H_{T}$, it will become possible to extract the tensor charge of the proton, which is a fundamental quantity and so far only poorly constrained.

The JML model, which turns out to reproduce available measurements of un-polartized electro-production cross-sections with a focus on $Q^2$ up to 5 GeV$^2$  and $W$ up to 4~GeV \cite{AHC08,PGG13}, provides a reasonable description of the sign and the shape of $\sigma_{LT'}/\sigma_{0}$ at low and medium $Q^{2}$ and $x_{B}$ values, but shows extrapolation problems for the highest $Q^{2}$ and $x_{B}$ bins for which no explicit tuning could be performed based on previous data. The predicted theoretical $\sigma_{LT'}/\sigma_{0}$ values also fall short by a factor of two on average to reproduce the experimental values. However, a better agreement can be observed in the region of the lowest investigated $Q^{2}$ values, while the difference increases for higher values of $Q^{2}$.
The observed effects may originate from missing ingredients in the model. For instance, only the dominant singular unitary part of the re-scattering integrals is taken into account, while the effect of the principal part may be significant in the interference with the pole amplitudes. However, the observed difference in magnitude may also reflect the smallness of the theoretical transverse amplitude, which also misses the experimental value by a factor two at lower $W$ \cite{JML20prog}.

The BSA measurement provides us with an access to the small box diagram contributions (either chiral odd GPD's or unitarity rescatterings), through their interference with the dominant pion pole amplitude. As $Q^{2}$ increases, the differences between the two theoretical approaches, as well as their departure from experiment, may tell us that they are used beyond their domain of validity: lack of other unitarity cuts in the Regge approach, energy  not large enough to safely replace the hadronic basis by the quark basis in the GPD approach. The experiment presented here calls for improvements of the models along these lines.


In summary, we have performed a multidimensional measurement of the structure function ratio $\sigma_{LT'}/\sigma_{0}$ for $\vec{e} p \to e^\prime n \pi^+$ at large photon virtuality, above the resonance region. The comparison in very forward kinematics showed that, especially, the magnitude of $\sigma_{LT'}/\sigma_{0}$ is overestimated in all $Q^{2}$ and $x_{B}$ bins by the most advanced GPD-based model \cite{GK09}, indicating that a new global fit for the dominating GPD $H_{T}$ becomes necessary to obtain a better fit for the dominant GPD $H_{T}$ and the directly related tensor charge of the proton.
Also the Regge-based JML model shows difficulties to fully reproduce the data and underestimates $\sigma_{LT'}/\sigma_{0}$ in the investigated  $Q^{2}$ and $x_{B}$ region. However, especially at low $Q^{2}$, the JML model shows a slightly better agreement than the GK model, while the situation changes for high $Q^{2}$ where the GPD-based model provides a better reproduction of the data, especially after the GPD $H_{T}$ is adjusted.


\input acknowledgement.tex


\end{document}

%% file: acknowledgement.tex
We acknowledge the outstanding efforts of the staff of the Accelerator and the Physics Divisions at Jefferson Lab in making this experiment possible.
We owe much gratitude to P. Kroll for many fruitful discussions concerning the interpretation of our results.
This work was supported in part by the U.S. Department of Energy, the National Science Foundation (NSF), the Italian Istituto Nazionale di Fisica Nucleare (INFN), the French Centre National de la Recherche Scientifique (CNRS), the French Commissariat pour l$^{\prime}$Energie Atomique, the UK Science and Technology Facilities Council, the National Research Foundation (NRF) of Korea, the Helmholtz-Forschungsakademie Hessen für FAIR (HFHF), the Ministry of Science and Higher Education of the Russian Federation and the National Natural Science Foundation of China. The Southeastern Universities Research Association (SURA) operates the Thomas Jefferson National Accelerator Facility for the U.S. Department of Energy under Contract No. DE-AC05-06OR23177.